\documentclass[10pt, conference]{IEEEtran}
\IEEEoverridecommandlockouts

\usepackage{amssymb,amsmath,color,graphicx, amsbsy}
\usepackage{subfigure}
\usepackage{url}
\usepackage{algorithm}
\usepackage{algorithmic}
\usepackage{amsfonts}
\usepackage{epstopdf}
\usepackage[utf8]{inputenc}
\usepackage{hyperref}
\usepackage{dblfloatfix}

\title{Torrit: A GUI-Based Power System Simulation Platform}
\author{Md Ashfaqur Rahman, Ph.D.}
\date{July 2020}

\begin{document}

\maketitle

\begin{abstract}
An adequate education on power system operations and controls requires a hands-on experience on a graphical user interface (GUI) based software. At present, most commercial software do not have free editions with high flexibility and most freeware do not have good interfaces. This paper introduces a GUI-based application called ``Torrit'' for executing operations of power systems, especially for transmission systems. It is written in Python for it's rapid development ability. Torrit's main window includes a single canvas with some standard graphical interactions like create, delete, copy, move, double click etc. The beta version of this application is the focus of this paper that allows executing, saving and re-opening a project in three different modes - per unit computations, power flow, and state estimation. However, it is still in a rudimentary stage and many extensions are planned for future to match the needs of both academia and industry.

\end{abstract}

\section{Introduction}
Simulation is one of the core processes of power system operations. It helps to take fast and accurate decisions, supports secure co-ordinations, and automates system processes. Power system operation and control algorithms are designed keeping the software implementation in mind.

The history of power system software is as old as the history of general purpose software. In fact, electronics-based automation prevailed the operation of power systems before the use of computers. Among leading software of today, PSS/E started its journey in 1972 \cite{psse}. PSCAD is serving the industry for over 40 years and it has the longest-lasting transient study tool \cite{pscad}. A good number of software emerged in the eighties including EasyPower (1984) \cite{easypower}, DIgSILENT (1985) \cite{digsilent}, and ETAP (1986) \cite{etap}. PowerWorld started its journey as a company in 1996 \cite{powerworld}. There are a number of open source software as well. Among those, MATPOWER, which was introduced in 1997, is one of the most popular tools for research \cite{matpower}. PSAT was introduced in 2005 \cite{psat}. A cloud-based solution is provided by a project called InterPSS \cite{interpss}.

Although there are a lot of applications available in the market, very few of those are free or inexpensive. The prices of professional software are out of reach of general students as well as of many researchers. In some cases, well-known software have student editions which tremendously limit the operations and do not make a proper sense of education. The open source applications, which are also free to use, are usually intended for only research, and they do not include a good interface to give the control-room feelings to the students. The details of calculations are either hidden from the user or substantially abstracted that makes it difficult to debug.

A software is needed for power systems which will have a professional edition for industry applications, and at the same time, it will have a free edition that will help develop future engineers. The concept is easily available in other fields of software industry. Nowadays, many software have a professional edition as well as a community edition. Some examples include Visual Studio (2013, 2015, 2017, 2019), PyCharm (2019), VSDC (v6.4.7.155) etc. Unlike the student versions of existing power system software, those allow a high flexibility with a lot of features that is enough for non-professional use. To sustain the profit of the company, professional editions are released.

This paper introduces a new Python based platform called  ``Torrit" that aims at serving both student and professional community. The use of Python has remained very limited in power systems so far. A few exceptions include PyPSA \cite{pypsa}, Pandapower \cite{pandapower}, DOME \cite{dome} etc.  Even though Python has many limitations, it is becoming more popular in the mainstream software industry. It is holding the mid position of fast-computing and low-development-time languages.

``Torrit" is a Bangla word that means ``Electricity". The application has a professional interface to create an aesthetic feelings to the user. Using this application is as easy as using other circuit analysis software like PSpice, Circuit maker etc. This helps new users to easily adapt the interface. The beta version only supports the development of single line diagram of the transmission system. Future releases may include distribution level design.

The main contributions of Torrit are as follows,
\begin{itemize}
    \item Torrit is aimed at serving the student as well as professional communities. It will have a significant amount of freely available features to support the learning of future generations.
    \item The main contribution of Torrit is it's easy-to-use graphical user interface (GUI). It is similar to other circuit simulators that helps new users.
    \item It is based on Python which is the most promising language of this decade. The extension of this application should be much simple compared to others developed with different languages.
\end{itemize}

The rest of the paper is organized as follows. The organization of Torrit is described in Sec. \ref{sec:organization}. The components and graphical interactions available in it are discussed in Sec. \ref{sec:comp}. The supported power system operations are described in Sec. \ref{sec:power}. A simplified class diagram for the project is briefly depicted in Sec. \ref{sec:class}. The plan for future expansions is analyzed in Sec. \ref{sec:plans}. The advantages of choosing Python and it's module \textit{tkinter} are discussed in Sec. \ref{sec:discussion}. The paper is concluded in Sec. \ref{sec:conclusion}.

\section{Organization of Torrit}\label{sec:organization}
The entire power system can be partitioned in three major levels - generation, transmission, and distribution. The analysis of each level differs significantly and those are usually simulated separately. For example, state estimation, contingency analysis, unit commitment, optimal power flow, transmission planning etc. are analyzed in transmission level, where fault analysis, arc flash etc. are taken under distribution level.

\subsection{User Interface for Transmission and Distribution Systems}
The distribution system is depicted with a three-phase diagram while the transmission system uses a simplified single-line diagram. There are a few other restrictions in transmission systems. A direct connection between two transmission lines is prohibited and in case two lines need connection, it should be done through a bus. However, it is allowed in the distribution system. 

In many cases, distribution system diagrams can also be abstracted with the single line diagram. For example, the connection between two lines can be abstracted with a three-port connecting component. The possibility will be analyzed in future. For now, Torrit is supporting only transmission system analysis with single line diagram.

\subsection{Windows in Torrit}
The user interacts with Torrit through windows. Windows are key elements of any GUI-based application. Torrit has three windows- main/root window, properties window, and calculation window.

\subsubsection{Root Window}
\begin{figure}[h]
    \centering
    \hspace*{0 cm}{\scalebox{0.2}{\includegraphics{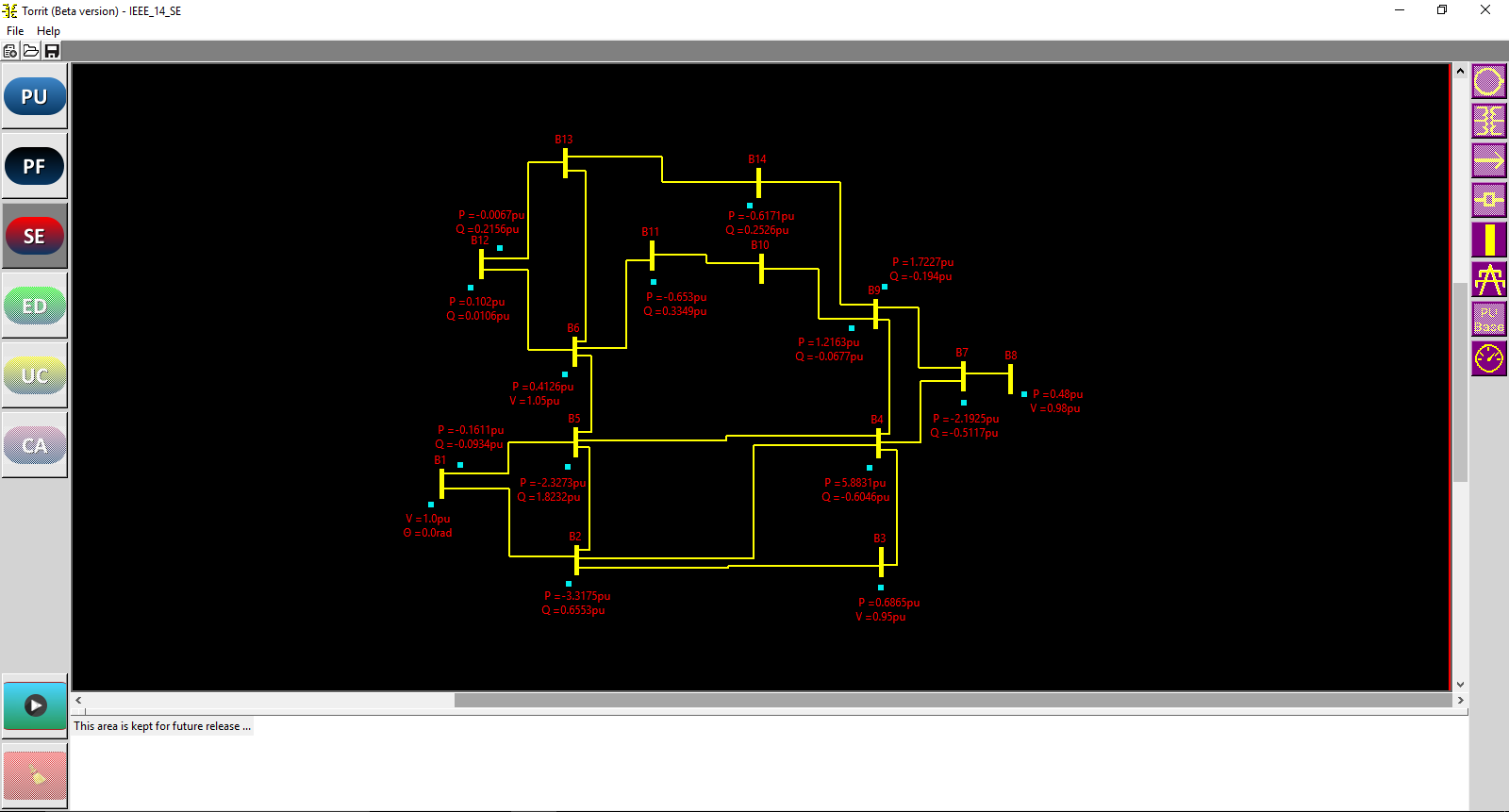}}}\vspace{0 cm}
    \caption{The main window of Torrit with the state estimation mode for IEEE 14-bus system.}
    \label{fig:main_window}
\end{figure}

Torrit has a master window that introduces the main features of the application. It contains a menubar with two sets of menus, and three toolbars on three sides of the window as shown in Fig. \ref{fig:main_window}. The left toolbar contains buttons for modes of operations such as state estimation, power flow etc. while the right one contains the components like generator, transformer, transmission line, bus-bar, load etc. The center part contains the canvas for drawing the network. There is no separate window for showing summarized results. It is shown on the canvas of the main window.

\subsubsection{Properties Window}
\begin{figure}[h]
    \centering
    \hspace*{0 cm}{\scalebox{0.4}{\includegraphics{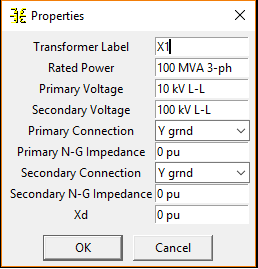}}}\vspace{0 cm}
    \caption{A sample properties window in Torrit for a transformer.}
    \label{fig:prop_window}
\end{figure}

Each component has its own properties which can be configured with related properties window. The properties window can be opened by double-clicking on the components. A sample window for a transformer is shown in Fig. \ref{fig:prop_window}. In the beta version, the properties are taken as a single string from the user that is parsed to separate different parts. For example, in this figure, the input for rated power is  `100 MVA 3-ph' which is separated to `100', `MVA', and `3-ph' for processing. It is considered as a weak method for input, and in later releases, separate controls are planned for each part of the properties.

\subsubsection{Calculation Window}
\begin{figure}[h]
    \centering
    \hspace*{0 cm}{\scalebox{0.3}{\includegraphics{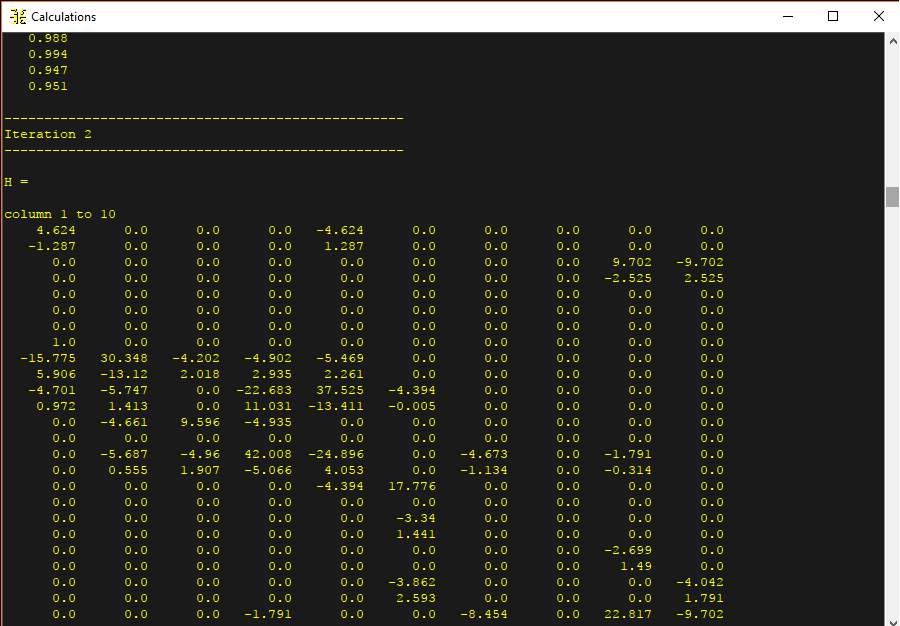}}}\vspace{0 cm}
    \caption{A sample calculation window of Torrit for the state estimation of IEEE 14-bus system.}
    \label{fig:calc_window}
\end{figure}

To make the calculations clear to the user, a calculation window is created for each operation. This is a unique feature of this application that helps the user understand the internal steps of each operation. It also works as a debug window that helps to boost the confidence of the user. A sample calculation window is shown in Fig. \ref{fig:calc_window}.

\section{Components and Graphical Interactions}\label{sec:comp}
\subsection{Components}
Torrit includes all basic components available in a transmission system. The components are contained in the right toolbar. A brief description of each one is given below,

\begin{itemize}
    \item Generator: A basic generator is included with only static properties. The dynamic properties will be included when dynamic analysis will be integrated in Torrit. It is available under per-unit calculations only.
    \item Transformer: It is a two-winding component that plays a crucial role in per-unit computations. The windings can be connected in delta or wye with or without impedance.
    \item Load: The loads can be configured with real and reactive power values. It can also be defined with R, L, and C values.
    \item Bus-bar: The bus-bar is considered as a passive component that does not have any system property. The visible length can be controlled using the `length' property.
    \item Transmission Line/Connecting Line: In Torrit, the transmission lines and connecting lines are handled together as a single component. When a transmission line has zero impedance, then it is considered as a connecting line.
    \item Meter: Meters are non-connectable objects and work based on the principle of minimum distance. If the closest object is a transmission line, the power of the meter will denote the power flow of the line. If the object is a bus, it will denote power injections to the bus. The symbol for meter is chosen to be a small rectangle as used in the standard practice.
\end{itemize}

\subsection{Graphical Interactions}
Torrit enables some standard graphical actions with the same keystrokes and mouse events used in other applications. Besides creating components on the canvas, it allows delete, rotate, move, copy, and double-click. However, it does not support selecting multiple objects, zoom, find, and undo options in the present release. A brief description of each interaction is given below,

\begin{itemize}
    \item Create: A component can be created by clicking on the component icon of the right toolbar and then another click on the canvas, as done in general circuit simulators. The creation of transmission line requires two valid connection ports which is described in Sec. \ref{subsec:trans}.
    \item Delete: Delete is performed on the selected component with the $<$Delete$>$ key of the keyboard. Torrit does not support any dangling line, and as a result, the delete event deletes any line connected to the deleted component to keep consistency.
    \item Rotate: It is associated with the keystroke $<$R$>$. It rotates all components by 90 degrees clockwise. There is one exception; the bus-bar cannot be rotated when connected with lines.
    \item Move: Movement is allowed for any component with mouse drag. In future releases, a fine movement is planned with arrow keys.
    \item Copy: The user can copy any object with $<$Ctrl+C$>$ and paste it in other part of the canvas. In current version, the user cannot select multiple objects at a time; so he cannot copy multiple items at a time as well.
    \item Double Click: It opens up the properties window for the selected component.
\end{itemize}

\subsection{Creation of Transmission Lines}\label{subsec:trans}
Transmission lines can be created by selecting the line from the right toolbar, and clicking on connection ports of any two components. For components other than bus-bars, there are one or two small rectangles that work as the connection ports. For the bus-bars, every enclosed point is considered as a connection port. Torrit does not allow dangling lines, i.e., every line that started from a valid connection port must end in another port. The lines are drawn in orthogonal pieces. When a component moves, the connected lines also move and the number of pieces change to three.

\subsection{Save and Open}
Like other standard software, Torrit allows the project to be saved and re-opened for further use. The project is saved in a single file with an extension \textit{.sld} which stands for single line diagram. Torrit uses two Python modules named \textit{simplejson}, and \textit{jsonpickle} to encode and decode necessary information to save and reproduce the project. A project-based saving option, where multiple subsystems can be saved in different files, is planned for future versions.

\section{Power System Operations}\label{sec:power}
\begin{figure}[h]
    \centering
    \hspace*{0 cm}{\scalebox{0.4}{\includegraphics{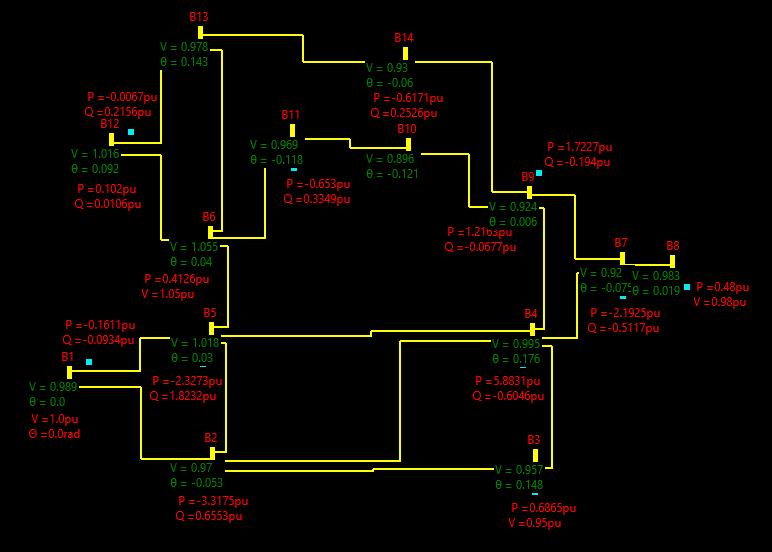}}}\vspace{0 cm}
    \caption{Visualisation of the estimated states in Torrit for IEEE 14-bus system.}
    \label{fig:result}
\end{figure}

In the beta version of Torrit, which was published in 2017 as a freeware, there are three operations available - per unit calculations, power flow, and state estimation. The operation modes are independent of each other and the output of one cannot be used by the other. For each mode, specific components are available related to it. The operation modes are selected first from the left toolbar of the main window. After creating or opening a project, the execution will show the results in both calculation and main window. An example is shown in Fig. \ref{fig:result}.

\begin{figure*}[b]
    \centering
    \hspace*{0 cm}{\scalebox{0.35}{\includegraphics{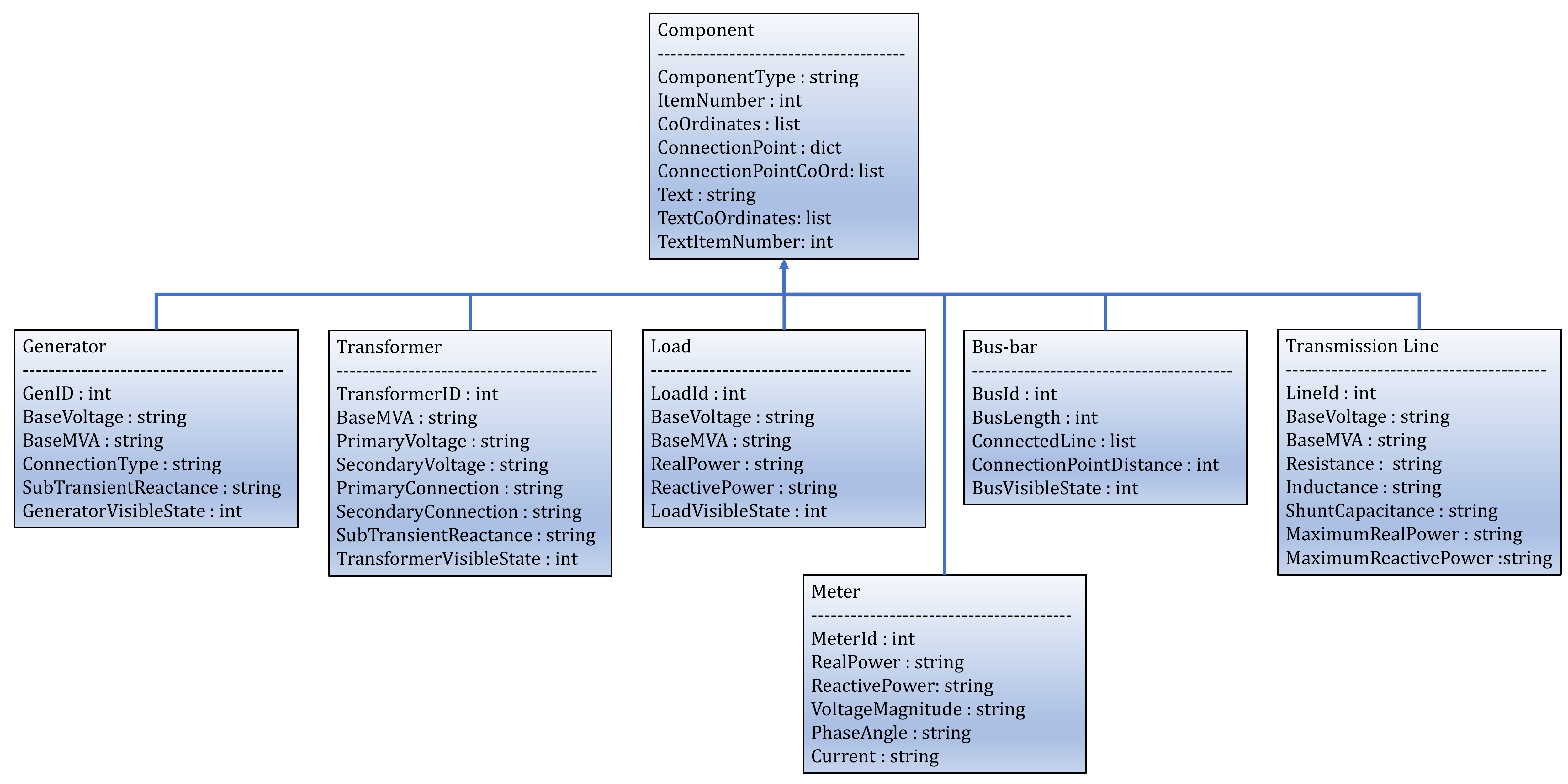}}}\vspace{0 cm}
    \caption{Component class diagram of Torrit with selected attributes.}
    \label{fig:class_diagram}
\end{figure*}

\subsection{Per-unit Computations}
Per unit calculations require a base power for the system and a base voltage for one part of the network. It is done by a special unit called $PU_{Base}$ which is also created like a component. It works on the principle of minimum distance. The base voltage of the closest component is assumed to be the base voltage provided by $PU_{Base}$.

\subsection{State Estimation}
State estimation is based on available measurements in the network. It has two options, weighted least squares (WLS) estimator, and fast decoupled state estimator (FD-SE). WLS estimator is the extended version of the Newton-Raphson method of power flow. FD-SE is the most popular estimator in the industry due to its fast execution.

\subsection{Power Flow}
For power flow, there are two standard methods, Gauss-Seidal method and Newton-Raphson method. Both are available in Torrit. For Gauss-Seidal method, the acceleration constant is taken as $1.6$. Ten iterations are shown for Gauss-Seidal method while five is for Newton-Raphson method. The numbers are usually enough for the convergence of the algorithms.

\section{Class Diagram}\label{sec:class}
A simplified class relation is used in the beta version of Torrit. As it does not allow multiple canvases, no class is needed for the canvas of the main window. Among many elements of the application, only the component classes require inheritance property as they share many common attributes like the co-ordinates and text of the item, co-ordinates of the text etc. The component classes with selected attributes are shown in Fig. \ref{fig:class_diagram}.

The classes could be organized in much sophisticated manner. For example, the visible states of generator, transformer, load and bus-bar could be taken under one child class, and transmission line and meter could be taken under another. However, to keep the things simple, those are not introduced in the beta version.

There is another discrete class to create the properties window of the components. It has only two attributes that contains the property names and corresponding values.

\section{Plans for Future Extensions}\label{sec:plans}
The beta version is an initial product and there are many possibilities for future extensions. The main challenge for the beta version was to create an interactive canvas that enables the drawing of single line diagrams. As the obstacle is crossed in this version, it is now ready to be extended.

\subsection{Multi-Canvas Projects}
The beta version supports a single canvas that limits the use of components. In future, a multi-canvas platform can be developed where each canvas will work as a subsystem. This will have the option to import subsystems developed somewhere else. It will serve three specific purposes, 
\begin{itemize}
    \item A transmission system may include tens of thousands of buses that may not fit in a limited canvas. A multi-canvas project can allow unlimited number of subsystems.
    \item A big system may need to be developed by multiple users. Importing subsystems will allow multiple users to work on different parts of the project simultaneously.
    \item The option of subsystems will reduce the complexity of developing large systems.
\end{itemize}

The option of multiple subsystem will create some challenges as well,
\begin{itemize}
    \item The connections between different subsystems will require a connecting component such as a bubble of PSpice. Each bubble must have exactly one matching bubble as zero match means a dangling line, and two or more matches means connecting three or more lines. The impedances of the lines connected to two conjugate bubbles need to be combined with predefined rules.
    \item Display of output will be significantly different than the existing one. It may be shown as a combined system on a single canvas in a separate window, or it can be shown on separate canvases in the main window (or in a separate window).
    \item The import of external subsystem may require checking for name conflicts and renaming of components as needed.
\end{itemize}

\subsection{Advanced Graphical Interactions}
A few interactions, which are available in standard applications, need to be added. Some of those include \textit{find, undo, zoom} etc. The \textit{find} event will look for a specific string in the visible properties of all components in the canvas. Multi-canvas system will create an overhead for \textit{find} event. The \textit{zoom} action will need to be applied on all canvases.

The \textit{undo} event will be implemented with a stack. Stack can be implemented with different data structures in Python. A convenient one is the \textit{list} that allows \textit{append (push)} and \textit{pop} methods.

\subsection{Co-ordination of Operations}
In the current version, only a few modes are available and those are not interchangeable as well. A user may want to create a single project and apply different power system operations on it. Enabling a universal project, that will allow a single diagram to be used by each operation, is not trivial. Each component has its own properties that may differ or overlap for different operations. Some properties may not be compatible with some operations. For example, the impedance of a generator may not be included in state estimation. A detailed analysis will be needed to separate the common and unique properties of each operation.

\section{Discussion}\label{sec:discussion}
\subsection{Python for Standalone Applications}
Python has grown as a scripting language and it was not used to develop standalone applications. This is mainly because it is an interpreted language that is translated during runtime. It makes the applications significantly slow compared to those developed with complied languages like C or C++. Any professional software should be fast and responsive. However, even after these drawbacks, Python is now being used to develop standalone as well as web applications. Some advantages of Python are discussed below,

\begin{itemize}
    \item The very first quality of Python is rapid development. The application can be developed within a short time with less effort. That, in turn, helps to keep the development cost low and makes the product cheaper.

    \item Python is a platform-independent language i.e. it can be run on different operating systems with little or no change in code. In Torrit, except some environment information like screen width and height, the code is free of any dependency on Windows operating system.
    
    \item It is a dynamically-typed language which matches the most widely used language for research, MATLAB. In fact, two modules named \textit{numpy} and \textit{scipy} are introduced in Python as alternatives to MATLAB. As a result, conversion from MATLAB to Python is much effective and research findings are easier to integrate.
    
    \item The codes written in C/C++ can be used by Python as external modules. Moreover, Python codes can be interpreted to C language for efficient execution using a static compiler named \textit{Cython}. However, it may create some issues in packaging the programs as a standalone application. 
    
    \item Python is taking a leading place in data science. As the applications of data science and machine learning in power systems are increasing significantly, the integration of data science tools like \textit{Tensorflow} or \textit{PyTorch} will be simpler.
    
    \item It is a growing language and it has support in almost every direction. Besides \textit{numpy}, and \textit{scipy} for scientific computations, it has modules like \textit{tkinter, PyQt} for GUI-based desktop applications, \textit{pymssql, pyodbc} for database support, \textit{django,flask} for web frameworks etc.
\end{itemize}

Even after all these advantages, the final plan for Torrit is to run the computation intensive part in C to fit industrial needs. In that case, the interface and related applications will run in Python and some matrices like branch, bus or generator matrices will be passed in C for computing. Parallel processing may also be considered in future.

\subsection{The choice of \textbf{tkinter}}
There are mainly two modules for creating GUI-based applications in Python. \textit{PyQt} is a developed module and it has an enriched library. On the other hand, \textit{tkinter} is a native one that comes with the standard Python distribution. It is lightweight, but lacks a number of standard widgets. To makeup the drawbacks, some advanced extensions like \textit{ttk} are introduced later. In Torrit, \textit{tkinter} is chosen to make the interface light. As there are a good number of interacting components on the canvas, it is expected to have a simple behavior to keep the application fast.

\section{Conclusion}\label{sec:conclusion}
Torrit is still in a primary stage in its evolution and a significant amount of improvements can be made in future. The main part of the project was to develop an interactive canvas with some basic graphical actions that enable the creation of single line diagrams. It also included a few power system operations to exemplify it's ability to support the needs.

A few graphical interactions need to be developed to enhance the experience of the user. Many power system operations need to be added as well. As a Python-based project, the development of new operations should be simple and fast. The integration of other tools should be simple as well.

\bibliographystyle{IEEEtran}
\bibliography{torrit.bib}

\section*{Appendix}
The beta version of Torrit can be downloaded from,
\\
\\
\url{https://drive.google.com/drive/folders/1mcuG7uymSfJZOjHXJA9Y68Z6nDsTZocC?usp=sharing}
\\
\\
Link for the project site,
\\
\\
\url{https://ashfaqeee.wixsite.com/torrit}

\section*{Biography}
Md Ashfaqur Rahman is currently serving as a Visiting Assistant Professor at Texas A\&M University - Kingsville. He received his doctorate in electrical engineering from Clemson University in 2017. Before that, he received his master’s degree from Texas Tech University in 2012. His research interests include cyber security of smart grid, power system state estimation, machine learning etc. He has published three journal articles and ten conference papers, and his publications have been cited more than 300 times. He has around seven years of teaching experience in undergraduate level in different universities of USA and Bangladesh. (email: \textit{ashfaq.eee@gmail.com})

\end{document}